# Electric Field-Induced Formation of a 2D Adatom Gas on Cryogenic Li Surfaces

Shyam Katnagallu[1,*], Huan Zhao[1,2], Se-Ho Kim[1,3], Baptiste Gault[1,4], Christoph Freysoldt[1,*], Jörg Neugebauer[1]

[1] Max Planck Institute for Sustainable Materials, Max-Planck-Straße 1, 40237 Düsseldorf, Germany.

[2] State Key Laboratory for Mechanical Behavior of Materials, Xi'an Jiaotong University, Xi'an, China

[3] Materials Science and Engineering, Korea University, Seoul 02841, Republic of Korea.

[4] Department of Materials, Imperial College London, Royal School of Mines, Exhibition Road, London, SW7 2AZ, UK.

*Corresponding authors: s.katnagallu@mpie.de; c.freysoldt@mpie.de;

## Abstract:

Intense electrostatic fields, such as those able to break bonds and cause field-ion emission, can fundamentally alter the behaviour of atoms at and on the surface. Using density functional theory (DFT) calculations on the Li (110) surface under high electrostatic fields, we identify a critical field strength at which surface atoms occupying a kink site become thermodynamically unstable against adatom formation. This mechanism leads to the formation of a highly concentrated two-dimensional (2D) adatom gas on the surface. Moreover, the applied field reverses the stability of preferred adsorption sites, enabling barrierless diffusion of lithium atoms even well below the threshold required for field evaporation. The here identified mechanisms offer a unified explanation for experimental observations in atom probe tomography and for understanding high electric field phenomena in systems such as battery interfaces and electrochemical environments.

## Letter:

Electric fields critically influence the adsorption and diffusion energies of adatoms, directly impacting surface dynamics and phase stability [1]. At sufficiently high electric fields, surface atoms may undergo field desorption, ionization and evaporation — phenomena that are the basis for advanced characterization techniques such as field ion microscopy (FIM) [2–4] and atom probe tomography (APT) [5,6]. These materials characterization techniques enable near-atomic resolution imaging and compositional analysis.

Previous studies have shown that electric fields can induce surface phase transitions in metals like lead (Pb) [7] and gold (Au) [8]. For Au, electric fields act as reversible triggers to induce surface roughening, as observed via in-situ transmission electron microscopy. Estimated electric fields of $10^9$ V/m were applied to Au nanocones, resulting in a disordered phase occurring at room temperature or higher. Ab initio molecular dynamics calculations, including electric fields, confirmed surface roughening in Au nanoparticles, although the underlying mechanism remained poorly understood. Electrostatic fields are also known to cause surface reconstructions, as theoretically analysed for platinum (Pt) and Au (110) surfaces [9], where reconstruction resulted from the interaction between surface charge and the work function.

In the present study, we will focus on the impact of high electric fields on Li surfaces. Li is widely used in modern technology due to its unique properties and wide-ranging applications involving high electric fields. As the active charge carrier in lithium-ion batteries, Li is critical for energy storage solutions, i.e. batteries that are both viable and scalable [10]. Li is also proving to be an unmatched element for most energy efficient production of ammonia [11]. These Li ion batteries power devices—from portable electronics to electric vehicles—are essential for integrating renewable energy sources into the grid. However, Li-ion batteries face significant challenges, such as dendrite formation during charging cycles. Dendrites are needle-like Li-rich deposits that can grow inside the battery, potentially causing short circuits and reducing battery life [12–14]. Additionally, the limited cycle life and safety concerns associated with dendrite-induced failures hinder the effectiveness of these batteries. The effect of electric fields is also crucial in understanding dendrite formation, where electric fields can reach $10^6$-$10^7$ V/m and even higher orders of magnitude in the early growth stage [15,16]. In addition, the low-melting point of Li makes it suitable for a liquid metal ion source in a focused ion beam (FIB) [17–20]. FIB systems are used for precise material removal and nanoscale imaging in semiconductor fabrication and materials science research. Despite its potential, using Li in FIB technology presents challenges, including the instability of Li ion sources due to its high reactivity and low melting point, and difficulty in controlling Li ion beams, which affects resolution and precision [21,22].

Motivated by these findings and the challenges in Li applications, we investigated the electrostatic field effects on Li surfaces by APT experiments and state-of-the-art ab initio simulations that directly incorporate the electrostatic field [23,24]. An important step in bringing theory and experiment together is the development of an approach to extend the powerful concept of DFT constructed surface phase diagrams to include the electric field as a state variable. Using this approach, we explore the high-field behaviour of Li surfaces and show that electrostatic fields can drive a phase transition from a conventional, largely adatom-free surface to a 2D adatom gas. The formation of such a dense adatom gas naturally explains the formation of Taylor cones in the APT experiments, a feature that has only been observed when probing liquids such as water [25]. By demonstrating that electrostatic fields effectively induce and regulate surface phenomena in Li, our results lay the ground for manipulating atomic-scale properties and advancing Li-based technologies. This encompasses not only more reliable APT analyses but also potential enhancements in battery performance and focused ion beam applications, ultimately underscoring the strategic importance of Li as a critical resource.

First, specimens for APT were prepared from a pure Li foil (Alfa Aesar, 99.9%) using a dual beam scanning electron microscope (SEM)/Xe – plasma focused ion beam (PFIB), FEI Helios PFIB, following the protocol proposed for pure Na (see Fig 1a and 1b) [26]. The prepared samples were transported to a CAMECA 5000 XS atom probe through an ultra-high vacuum (UHV) suitcase [27]. APT measurements were conducted at a base temperature of 60K, using laser mode with an energy of 15pJ and a frequency of 100kHZ at a detection rate of 0.4%. The detector event histogram, bottom inset in Figure 1c, shows concentrated spots of ion hits, which when reconstructed look like the elongated plumes seen in Figure 1c, resembling those reported in Li-containing oxides [28–30]. This is in stark contrast to "normal" APT datasets, which typically show uniform field evaporation [31–35]. There is no measurable influence of this field evaporation behaviour on the time-of-flight measurements. The mass-to-charge-ratio can hence be properly calculated, Figure 1d, and the two isotopes of Li are correctly identified.

The formation of hotspots on the detector has previously been associated to the formation of so-called Taylor cones [25]. These conical protrusions can form on liquid surfaces under strong electric fields, and the locally higher curvature at the apex of the cone enhances the electric field, facilitating field evaporation. Similar patterns have been reported in the APT analysis of water [25]. By virtue of field evaporation these sharp features smoothen out, and field evaporation continues by formation of new cones at a different location. As the reconstruction is built sequentially from the ion impact positions [32], this leads to the appearance of individual plumes, that each correspond to the formation and exhaustion of a cone.

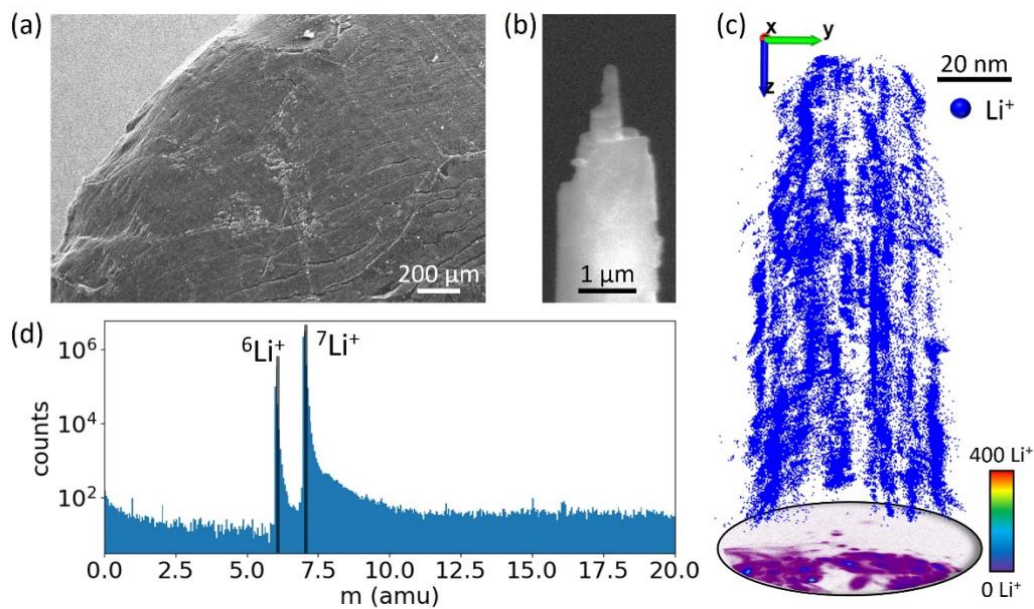

*Figure 1: (a) SEM image of a sliced Li bulk. (b) APT specimen of cryo-FIB prepared pure Li sample. (c) Reconstructed 3D atom map with a detector contour map (0 to 400 ion counts) map in a bottom. (d) The corresponding mass spectrum with two naturally occurring Lithium isotopes, with a measured isotope ratio of 13.2 (theoretical 12.2).*

Based on these experimental observations, we hypothesize that the Li surface undergoes a dramatic change from its stable crystal structure under these high electrostatic fields. To investigate this hypothesis, we conducted systematic density functional theory (DFT) calculations that explicitly include high electrostatic fields [23] to investigate the behaviour of Li metal surfaces. On the surface, steps and kinks are the relevant sites for field ion imaging and field evaporation due to their lower coordination and thus weaker bonding. To represent such features using periodic slab models that are computable within DFT, carefully chosen high-index surface orientations are employed. To set up the simulations, we designed a tool to generate metallic slabs of a high Miller index [36] with selected terrace, length, and direction of steps and kinks, inspired by the work by Van Hove and Somorjai [37]. This tool and all the DFT calculations we discuss later are set up and analysed in pyiron [38], which is an integrated development environment for atomistic simulations. We created a bcc Li (110) surface with 65 atoms. To apply an electrostatic field on one side of the slab, we use the generalized dipole correction implemented in the DFT library S/PHI/nX [23,39]. For the calculations, we use the projector augmented wave pseudopotentials taken from the VASP library and the Perdew–Burke-Ernzerhof

(PBE) [40] exchange-correlation functional. We ensured a careful convergence with respect to the **k** points and the energy cutoff for plane waves to yield an accuracy of $10^{-3}$ eV in the relevant energies. A **k**-point sampling of 6x6x1 for (110) cells and equivalent k-density sampling for (952) with a plane-wave energy cutoff of 550 eV was used.

The electrostatic field provides a degree of freedom for phase stability and has two dominant effects when applied at sufficient strength. First is the atomic scale charge (electronic) redistribution, which either softens or strengthens some bonds. Second is the induction of a tensile stress on the slab, called Maxwell stress that requires an additional convergence criterion—namely, the total thickness of the slab (≈18 Å) and the height of the fixed layers (7.5 Å) in the slab—to yield converged interlayer relaxations to within 0.1%. Additional care must be taken for electronic occupancy smearing methods and widths used, as the charge distributions under these high electrostatic fields proved to be sensitive to them. We used Fermi-Dirac smearing with first-order entropy corrections with a width of 0.1 eV.

To understand how the electrostatic field impacts the surface phases, we chose to compute the adsorption surface phase diagram for the (110) surface of Li. For bcc Li, (110) is a close-packed surface and has only slightly higher surface energy than the most stable (100) surface under field free conditions [41,42]. The self-adsorption energy of Li-on-Li is calculated in the presence of an electric field $\mathcal{E}$ as

$$E_{ad}^{Li}(\mathcal{E}) = E_{tot}^{slab+Li^{ad}}(\mathcal{E}) - E_{tot}^{slab}(\mathcal{E}) - E_{ref}^{Li}(\mathcal{E}),$$

(1)

where $E_{tot}^{slab+Li^{ad}}$ is the total DFT energy of a metal surface with Li adatom, $E_{tot}^{slab}$ is the total energy of the clean (110) surface, and $E_{ref}^{Li}$ is the reference state to define the reservoir of Li adatom. On this particular Li surface, we see two possible adsorption sites, namely on-top and bridge. All these energies have been calculated as a function of electrostatic field, to get the adsorption surface phase diagram shown in Figure 2. We note, that in contrast to common surface phase diagrams the state variable is not chemical potential or temperature but electric field, allowing us to directly identify the thermodynamically preferred surface phase for a given electric field.

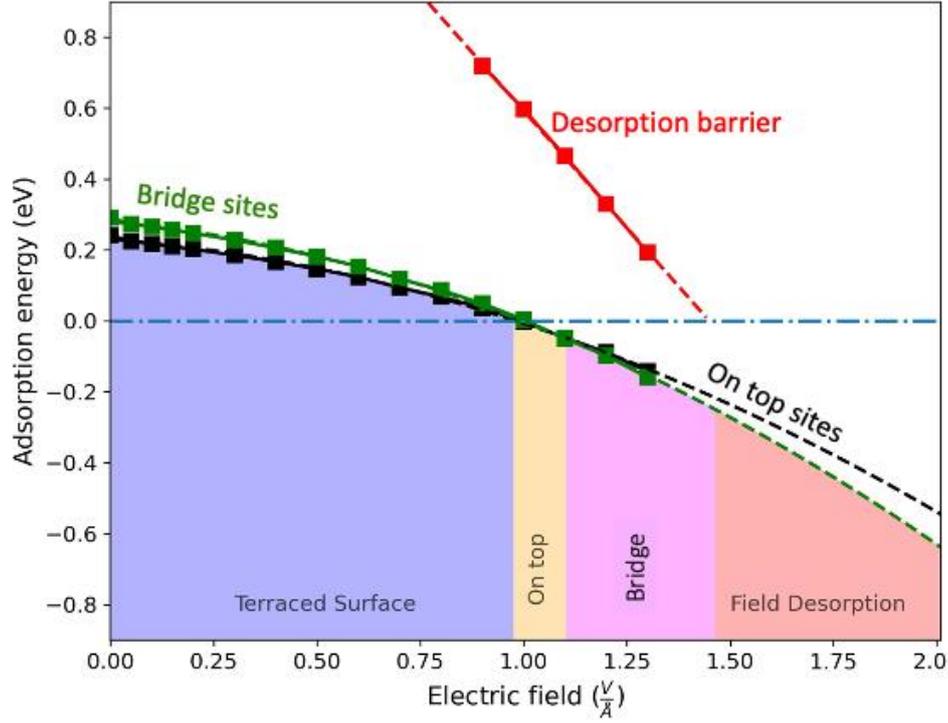

*Figure 2: The adsorption surface phase diagram of Li calculated as a function of the applied electrostatic field. The adsorption energies for the on-top (black) and bridge (green) sites are plotted as a function of electrostatic field. The field desorption barrier (for on-top) as a function of field is also plotted (red).*

The choice of reference is important for calculation of such a diagram. We considered the kink sites of a stepped Li surface as the reservoir for Li adatoms. These kink sites are a consequence of the hemi-spherical end-form of APT specimens. Kink sites represent extended reservoirs because removing an atom from such a site merely shifts the kink position along the step. Field evaporation usually occurs mainly from such kink sites due to the amplification of local electric fields [43–45] compared to terraces and straight steps. For calculations, we chose a (952) high index surface as the computationally feasible stepped-kink surface, which has a (110) type terrace with steps along (100) direction and kinks along (111). Hence, the $E_{ref}^{Li}$ is calculated as

$$E_{ref}^{Li}(\mathcal{E}) = E_b^{Li-kink}(\mathcal{E}) = E_{tot}^{slab+kink}(\mathcal{E}) - E_{tot}^{slab-kink}(\mathcal{E})$$

*(2)*

where, $E_b^{Li-kink}(\mathcal{E})$ is the binding energy of Li kink atom as function of electrostatic field, which is the difference of total energy of slab containing a Li kink site ($E_{tot}^{slab+kink}(\mathcal{E})$) and total energy of the slab after removing the kink atom ($E_{tot}^{slab-kink}(\mathcal{E})$). Using this reference, the adsorption energies of Li on two sites i.e., on-top and bridge sites are calculated at different electrostatic fields. In the field-free case, the kink-site reference is identical to the chemical potential of the bulk because the surfaces before and after removal are translationally equivalent. In the presence of the field, there is a small difference due to the extra field-induced energy in the volume of the atom. For periodic surfaces considered in our computations, it can be expressed directly as

$$\Delta E = \frac{1}{2}\epsilon_0 \mathcal{E}^2 \Delta\Omega = \frac{1}{2} Q\mathcal{E}\Delta z \,,$$

where $\epsilon_0$ is the dielectric constant of the vacuum, $\Delta\Omega$ the volume per atom of bulk Li, $Q$ the charge per surface unit cell and $\Delta z$ the displacement of the periodic surface along the surface normal when the atom is removed. The equivalence of the two expressions is a direct consequence of the relation of surface charge density and electric field,

$$\frac{Q}{A} = \epsilon_0 \mathcal{E}.$$

In field-free cases, the on-top site is energetically favored over the bridge site for Li by about 50 meV. This agrees with an earlier DFT study on Li adsorption energies [41]. This contrasts with the common adsorption site preference in other metals, where either bridge or hollow sites are preferred due to their higher coordination number. The application of electrostatic field shows a further stabilization of adsorption sites [1] compared to the kink-site reference. This is expected as the additional charge due to electrostatic field can be localized on a single atom closer to the counter-electrode, leading to the stabilization of the adsorption sites.

Interestingly, at fields above ~1 V/Å, both adsorption sites are thermodynamically preferred over the kink site, as the adsorption energies becomes negative. This implies a spontaneous dissolution of kink sites, to move to the adsorption sites, preferentially to the on-top sites. Thus, the region marked by the blue-coloured "terraced surface", is the most stable surface configuration for fields up to ~1 V/Å. Further increase in the electrostatic field lowers the adsorption energies. Up to ~1.1 V/Å, the on-top adsorption site continues to be more stable than the bridge site. However, the energetical differences among them are very small compared to the thermal energy at 60K, which is ~5meV. At electrostatic fields greater than 1.1 V/Å, the bridge site adsorption energy is lower than the on-top adsorption energy, leading to a switch in the favoured adsorption site from on-top to bridge site. These regions are marked in the adsorption-surface phase diagram where the on-top and bridge site adsorption regimes are marked in yellow and pink respectively. The concentration of the adsorbed Li atoms remains finite, however, due to a repulsive interaction between them, which would raise the adsorption energy and thus require higher fields to become thermodynamically competitive. To verify this, we calculated the surface phase diagram for adsorption as a function of both electric field and the coverage of Li adatoms, see supplementary figure S2. These calculations show the repulsive interactions between the adatoms increase the "critical" electric fields for thermodynamically spontaneous dissolution of kinks with increasing adatom coverage. At higher coverages the "critical" field becomes comparable to the field desorption strength.

Under such high electrostatic fields, field-induced desorption of the adatoms becomes prevalent. Usually, the adatoms have higher propensity to get field desorbed than atoms embedded into the surface, so it is relevant to evaluate the field desorption barriers for the adatom. For this we have calculated the potential energy profile as seen by an on-top adatom, as it is gradually moved away from the (110) surface. On this path, the atom is gradually acquiring a positive net charge [23,24,46]. In other words, the adatom gradually turns into an ion, also known as the charge-draining mechanism of field desorption [47]. At a given sub-critical field, this results in a 1D potential energy curve with a barrier that must be overcome by the adatom to become a free ion. Figure 3(a) shows these field desorption curves for the on-top adatom for fields ranging from 0.9 V/Å to 1.3 V/Å, with a significant barrier for field desorption. Even at a field strength of 1.3 V/Å, the barrier is estimated to be 0.15eV, positioned at 2.3 Å away from the adatom site. Field desorption is therefore unlikely under these conditions, and stable adatoms should be seen on the Li surface even at these high electrostatic fields. The field desorption barriers are extracted from these curves and are plotted in Figure 2, to show the relevant electrostatic field strengths where field desorption is to be expected. A linear extrapolation to the "zero" barrier is found to be at ~1.43 V/Å. Thus, the on-top site for adsorption continues to be stable till ~1.43 V/Å, after which the adatom is expected to field desorb. The field desorption barriers for bridge site adatom show a very similar profile (see supplementary, figure. S1).

Another relevant information to be considered is the field evaporation curves for kink site Li atoms as these kink sites are shown to dissolve and move to adatom sites from our calculations shown in Figure 2. The field evaporation curves for a kink atom on the (952) Lithium surface for fields ranging from 0.8 V/Å to 1.5 V/Å are shown in Figure 3(b). The field evaporation curves for the kink atoms are computed similarly as the field desorption curves. A barrier of 0.13 eV is still seen at 1.3 V/Å indicating that the field evaporation is less likely to occur. We also observed the roll over motion of kink Li atom on to the terrace in some cases. This also suggests that the kink atoms would either dissolve and diffuse on the terrace or can become adatoms on the terrace due to roll over motion. These calculations indicate that the dissolution of kink atoms to adatom sites is a very likely phenomenon to occur before either field desorption or field evaporation can occur.

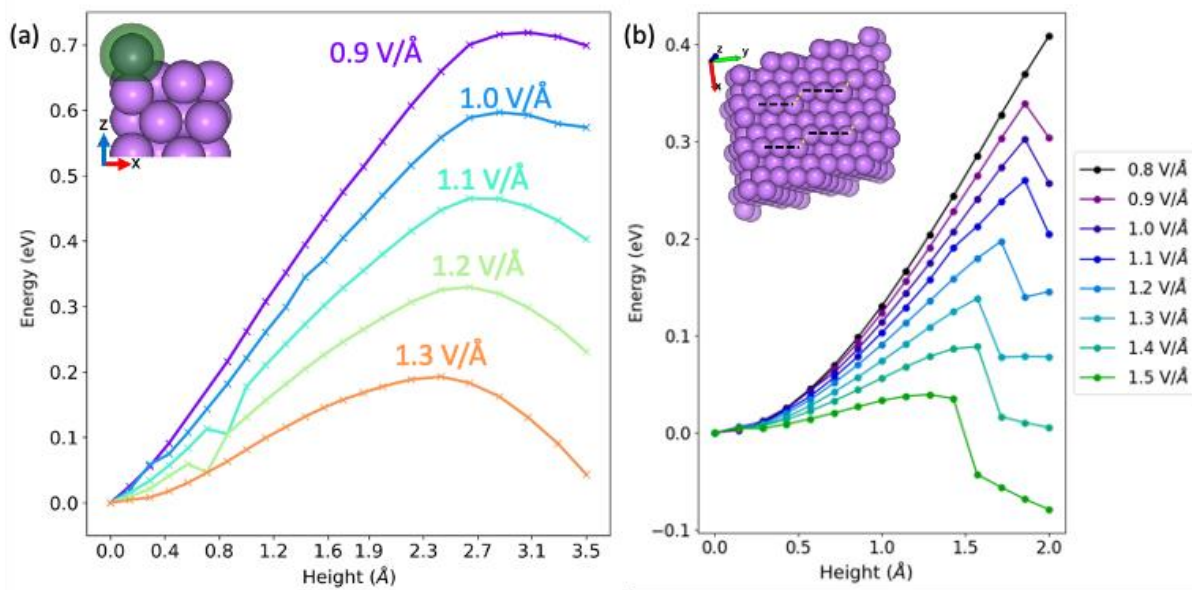

*Figure 3: (a) Field desorption curves for on-top adatom on Lithium (110) surface showing the presence of barrier. (b) Field evaporation curves for the kink atom evaporation from a (952) stepped Lithium surface. The insets show the corresponding surface configurations.*

To understand these peculiar effects of the electrostatic field on Li, we looked at the interatomic distances between the adsorbed Li atom and its first nearest neighbour. Li, when adsorbed on an on-top site, tends to form a dimer like bond with the surface atom, with an interatomic distance of 2.65 Å. We also performed a DFT calculation for the $Li_2$ molecule and obtained a bond distance of 2.7 Å, which is comparable to the observed first neighbour distance for on-top adsorption site. We therefore conclude that the on-top lithium atom forms a covalent bond with the surface atom. Such unique bonding character for the on-top site also leads to a reduced height of the on-top adatom compared to the bridge site adatom. Figure 4 (a) shows the heights of atoms for on-top and bridge configurations, respectively. Due to the difference in the heights of the adsorbed atoms and the possible formation of "dimer"-like bonds for the on-top site, the on-top atom can develop a smaller positive charge. This picture is confirmed by the Hirshfeld charge [48] on the adatoms, too. Figure 4(b) shows the Hirshfeld charge of the adatom at 1.2 V/Å field for either configuration. The on-top site develops an overall lower charge of 0.37$e$ than the bridge site atom's charge of 0.39$e$.

In Figure 2, we also saw that the adsorption energies of both sites are very similar near the critical field, where kinks can spontaneously dissolve. To understand the surface diffusion behaviour of the adatom, we calculated the energy profile along the green path from the bridge site to the on-top site, shown in the inset of Figure 4 (c). The energy profiles are referenced with respect to the bridge site and are calculated for various magnitudes of electrostatic fields. As seen previously when the electrostatic field is absent the on-top site is the stable site compared to the bridge site. The relative stability of the on-top site changes with the introduction of the electrostatic field. Above 1.1 V/Å, the bridge site becomes relatively more stable than the on-top site. Apart from that, in figure 4(c) it is evident that the energy profile becomes almost flat at 1.1 V/Å and further increase of the electrostatic field destabilizes the on-top site. The almost flat energy surface indicates a barrier less surface diffusion of Li adatoms could occur on the Li surface. The electrostatic field thus affects not only the relative stabilities of the adsorbed sites, but also the surface diffusion barriers.

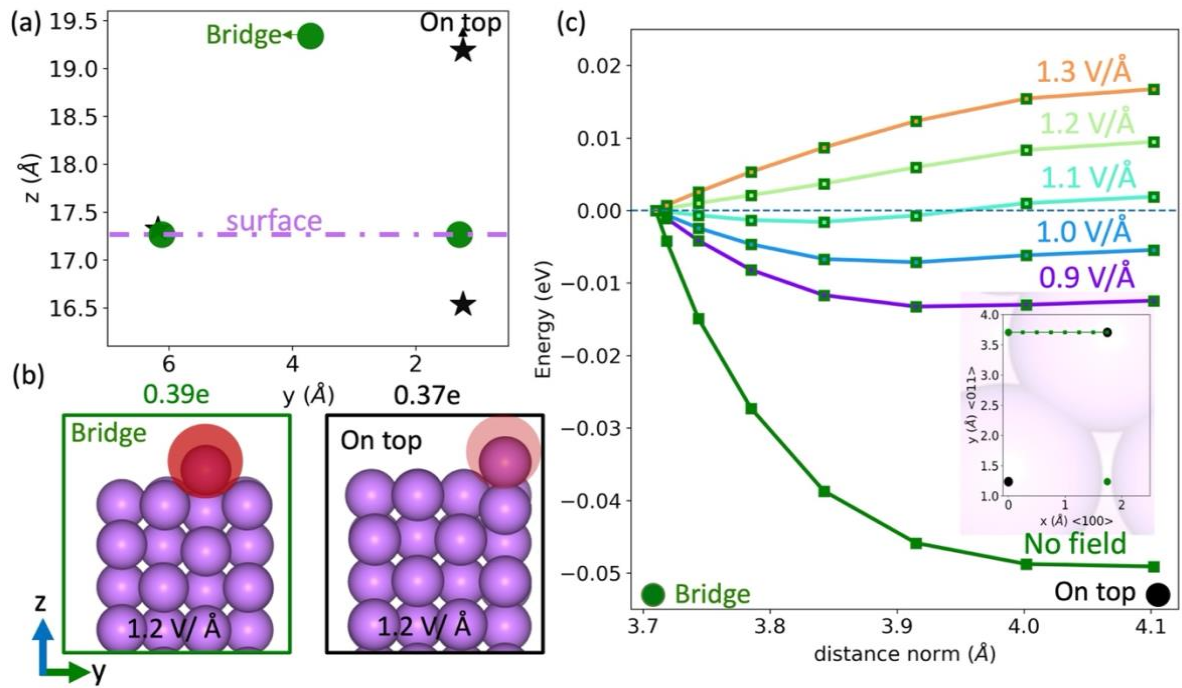

Figure 4: (a) z-y positions of the atoms in bridge configuration (in green, circles) and on-top configuration (in black, stars), indicating that bridge site protrudes slightly outwards compared to the on-top atom. (b) The Hirschfeld charge acquired by the bridge atom and on-top atom at 1.2 V/ Å field. (c) The energy profile along the green path from bridge site to on-top shown in the inset, for the various fields. The switch in the preferred adsorption site as a function of field and the reduction in diffusion barrier is seen.

The electrostatic field is known to reduce the surface diffusion barrier [1]. In addition, the field evaporation and desorption strength being greater than the field at which the switch in the preferred adsorption sites happens, leads to a very mobile layer on the surface of Li under electrostatic field. This allows for the Li surface to undergo a phase transformation from a kink/stepped surface to a 2D adatom gas phase (due to no surface diffusion barriers), which is stable.

In summary, by extending the concept of surface phase diagrams to include the electric field as a state variable, we have demonstrated the huge impact that electric fields, even below the field desorption limit, can have on surface stability and kinetics. In particular, for Li as a technologically highly relevant prototype material, we have identified regions in the phase diagram where the crystalline surface becomes thermodynamically unstable against the formation of a gas/liquid-like surface state, a state that would not be expected to form at the low temperatures encountered in a cryogenic environment. The high sensitivity of surface states and diffusion barriers to electric fields provides an important means of actively manipulating them using an easy-to-control quantity such as the electric field. Such phase diagrams are also expected to be very useful in identifying conditions where selected phases are absent, e.g. where such phases would lead to degradation or dendrite formation. Finally, we note that the approach outlined here to construct surface phase diagrams with the electric field as a state parameter from DFT surface calculations, where the electric field has been explicitly included, is general and can be easily applied to any system and surface structure.

**References:**

[1]     J. Neugebauer and M. Scheffler, Theory of adsorption and desorption in high electric fields, Surf Sci **287–288**, 572 (1993).

[2]     E. W. Müller and K. Bahadur, Field Ionization Of Gases At A Metal Surface And The Resolution Of The Field Ion Microscope, Physical Review **102**, 624 (1956).

[3]     F. Vurpillot, F. Danoix, M. Gilbert, S. Koelling, M. Dagan, and D. N. Seidman, True Atomic-Scale Imaging in Three Dimensions: A Review of the Rebirth of Field-Ion Microscopy, Microscopy and Microanalysis **23**, 210 (2017).

[4]     F. Danoix and F. Vurpillot, Basics of Field Ion Microscopy, Atom Probe Tomography: Put Theory Into Practice 73 (2016).

[5]     E. W. Müller, J. A. Panitz, and S. B. McLane, The {Atom-Probe} Field Ion Microscope, Review of Scientific Instruments **39**, 83 (1968).

[6]     B. Gault, A. Chiaramonti, O. Cojocaru-Mirédin, P. Stender, R. Dubosq, C. Freysoldt, S. K. Makineni, T. Li, M. Moody, and J. M. Cairney, Atom probe tomography, Nature Reviews Methods Primers 2021 1:1 **1**, 1 (2021).

[7]     B. Pluis, A. W. D. Van Der Gon, J. W. M. Frenken, and J. F. Van Der Veen, Crystal-Face Dependence of Surface Melting, Phys Rev Lett **59**, 2678 (1987).

[8]     L. De Knoop, M. Juhani Kuisma, J. Löfgren, K. Lodewijks, M. Thuvander, P. Erhart, A. Dmitriev, and E. Olsson, Electric-field-controlled reversible order-disorder switching of a metal tip surface, Phys Rev Mater **2**, 85006 (2018).

[9]     A. Y. Lozovoi and A. Alavi, Reconstruction of charged surfaces: General trends and a case study of Pt(110) and Au(110), Phys Rev B **68**, 245416 (2003).

[10]    T. Kim, W. Song, D. Y. Son, L. K. Ono, and Y. Qi, Lithium-ion batteries: outlook on present, future, and hybridized technologies, J Mater Chem A Mater **7**, 2942 (2019).

[11]    O. Westhead, R. Jervis, and I. E. L. Stephens, Is lithium the key for nitrogen electroreduction?, Science (1979) **372**, 1149 (2021).

[12]  A. J. Sanchez, E. Kazyak, Y. Chen, K. H. Chen, E. R. Pattison, and N. P. Dasgupta, Plan-View Operando Video Microscopy of Li Metal Anodes: Identifying the Coupled Relationships among Nucleation, Morphology, and Reversibility, ACS Energy Lett **5**, 994 (2020).

[13]  C. Niu, H. Lee, S. Chen, Q. Li, J. Du, W. Xu, J. G. Zhang, M. S. Whittingham, J. Xiao, and J. Liu, High-energy lithium metal pouch cells with limited anode swelling and long stable cycles, Nature Energy 2019 4:7 **4**, 551 (2019).

[14]  Y. Lu, Z. Tu, and L. A. Archer, Stable lithium electrodeposition in liquid and nanoporous solid electrolytes, Nature Materials 2014 13:10 **13**, 961 (2014).

[15]  E. Santos and W. Schmickler, The Crucial Role of Local Excess Charges in Dendrite Growth on Lithium Electrodes, Angewandte Chemie International Edition **60**, 5876 (2021).

[16]  P. Barai, K. Higa, Y. Dai, al -, J. Wang, H. Lin, and S. Passerini -, Spatially Resolved Growth Mechanisms of a Lithium Dendrite Population, J Electrochem Soc **170**, 030533 (2023).

[17]  P. M. Read, J. T. Maskrey, and G. D. Alton, A lithium liquid metal ion source suitable for high voltage terminal applications, Review of Scientific Instruments **61**, 502 (1998).

[18]  E. Hesse, F. K. Naehring, and J. Teichert, A lithium liquid metal ion source with a narrow angle emission for writing beam lithography, Microelectron Eng **23**, 111 (1994).

[19]  L. Bischoff, P. Mazarov, L. Bruchhaus, and J. Gierak, Liquid metal alloy ion sources—An alternative for focussed ion beam technology, Appl Phys Rev **3**, 021101 (2016).

[20]  S. Sivaprakash, S. B. Majumder -, P. Liu, J. Wang, J. Hicks-Garner, al -, F. Holtstiege, A. Wilken, L. Bischoff, and C. Akhmadaliev, An alloy liquid metal ion source for lithium, J Phys D Appl Phys **41**, 052001 (2008).

[21]  E. Hesse, L. Bischoff, and J. Teichert, Angular distribution and energy spread of a lithium liquid metal ion source, J Phys D Appl Phys **28**, 1707 (1995).

[22]  E. Hesse and F. K. Naehring, Narrow angle emission from a lithium liquid metal ion source, J Phys D Appl Phys **26**, 717 (1993).

[23]  C. Freysoldt, A. Mishra, M. Ashton, and J. Neugebauer, Generalized dipole correction for charged surfaces in the repeated-slab approach, Phys Rev B **102**, 45403 (2020).

[24]  M. Ashton, A. Mishra, J. Neugebauer, and C. Freysoldt, Ab initio Description of Bond Breaking in Large Electric Fields, Phys Rev Lett **124**, (2020).

[25]  T. Geoffrey Ingram, Disintegration of water drops in an electric field, Proc R Soc Lond A Math Phys Sci **280**, 383 (1964).

[26]  K. Thompson, D. Lawrence, D. J. Larson, J. D. Olson, T. F. Kelly, and B. Gorman, In situ site-specific specimen preparation for atom probe tomography., Ultramicroscopy **107**, 131 (2007).

[27]  L. T. Stephenson et al., The Laplace Project: An integrated suite for preparing and transferring atom probe samples under cryogenic and UHV conditions, PLoS One **13**, e0209211 (2018).

[28] B. Pfeiffer, J. Maier, J. Arlt, and C. Nowak, In Situ Atom Probe Deintercalation of Lithium-Manganese-Oxide, Microscopy and Microanalysis **23**, 314 (2017).

[29] S. H. Kim, S. Antonov, X. Zhou, L. T. Stephenson, C. Jung, A. A. El-Zoka, D. K. Schreiber, M. Conroy, and B. Gault, Atom probe analysis of electrode materials for Li-ion batteries: challenges and ways forward, J Mater Chem A Mater **10**, 4926 (2022).

[30] M. P. Singh, E. V. Woods, S. H. Kim, C. Jung, L. S. Aota, and B. Gault, Facilitating the Systematic Nanoscale Study of Battery Materials by Atom Probe Tomography through in-situ Metal Coating, Batter Supercaps **7**, e202300403 (2024).

[31] F. Vurpillot and C. Oberdorfer, Modeling Atom Probe Tomography: A review, Ultramicroscopy **159**, 202 (2015).

[32] D. J. Larson, B. Gault, B. P. Geiser, F. De Geuser, and F. Vurpillot, Atom probe tomography spatial reconstruction: Status and directions, Curr Opin Solid State Mater Sci **17**, 236 (2013).

[33] B. Gault, S. T. Loi, V. J. Araullo-Peters, L. T. Stephenson, M. P. Moody, S. L. Shrestha, R. K. W. Marceau, L. Yao, J. M. Cairney, and S. P. Ringer, Dynamic reconstruction for atom probe tomography, Ultramicroscopy **111**, 1619 (2011).

[34] F. Vurpillot, M. Gruber, G. Da Costa, I. Martin, L. Renaud, and a. Bostel, Pragmatic reconstruction methods in atom probe tomography, Ultramicroscopy **111**, 1286 (2011).

[35] B. Geiser, D. Larson, E. Oltman, S. Gerstl, D. Reinhard, T. Kelly, and T. Prosa, Wide-Field-of-View Atom Probe Reconstruction, Microscopy and Microanalysis **15**, 292 (2009).

[36] S. Katnagallu, C. Freysoldt, B. Gault, and J. Neugebauer, Ab initio vacancy formation energies and kinetics at metal surfaces under high electric field, Phys Rev B **107**, 41406 (2023).

[37] M. A. Van Hove and G. A. Somorjai, A new microfacet notation for high-Miller-index surfaces of cubic materials with terrace, step and kink structures, Surf Sci **92**, 489 (1980).

[38] J. Janssen, S. Surendralal, Y. Lysogorskiy, M. Todorova, T. Hickel, R. Drautz, and J. Neugebauer, pyiron: An integrated development environment for computational materials science, Comput Mater Sci **163**, 24 (2019).

[39] S. Boeck, C. Freysoldt, A. Dick, L. Ismer, and J. Neugebauer, The object-oriented DFT program library S/PHI/nX, Comput Phys Commun **182**, 543 (2011).

[40] J. P. Perdew, K. Burke, and M. Ernzerhof, Generalized Gradient Approximation Made Simple, Phys Rev Lett **77**, 3865 (1996).

[41] D. Gaissmaier, D. Fantauzzi, and T. Jacob, First principles studies of self-diffusion processes on metallic lithium surfaces, J Chem Phys **150**, 041723 (2018).

[42] K. Doll, N. M. Harrison, and V. R. Saunders, A density functional study of lithium bulk and surfaces, Journal of Physics: Condensed Matter **11**, 5007 (1999).

[43] Yu. Suchorski, N. Ernst, W. a. Schmidt, V. K. Medvedev, H. J. Kreuzer, and R. L. C. Wang, Field desorption and field evaporation of metals, Prog Surf Sci **53**, 135 (1996).

[44] I. M. Mikhailovskij, G. D. W. Smith, N. Wanderka, and T. I. Mazilova, Non-kinkwise field evaporation and kink relaxation on stepped W(1 1 2) surface, Ultramicroscopy **95**, 157 (2003).

[45] T. T. Tsong, *Atom-Probe Field Ion Microscopy: Field Ion Emission, and Surfaces and Interfaces at Atomic Resolution* (Cambridge University Press, 2005).

[46] C. G. Sánchez, A. Y. Lozovoi, & A. Alavi, C. G. Sa´nchez, S. Sa´nchez, and A. Alavi, Field-evaporation from first-principles, Mol Phys **102**, 1045 (2004).

[47] R. Gomer and L. W. Swanson, Theory of Field Desorption, J Chem Phys **38**, 1613 (1963).

[48] F. L. Hirshfeld, Bonded-atom fragments for describing molecular charge densities, Theor Chim Acta **44**, 129 (1977).

# Supplementary Information for Electric Field-Induced Cryogenic Formation of a 2D Adatom Gas on Li Surfaces

Shyam Katnagallu[1,*], Huan Zhao[1,2], Se-Ho Kim[1,3], Baptiste Gault[1,4], Christoph Freysoldt[1,*], Jörg Neugebauer[1]

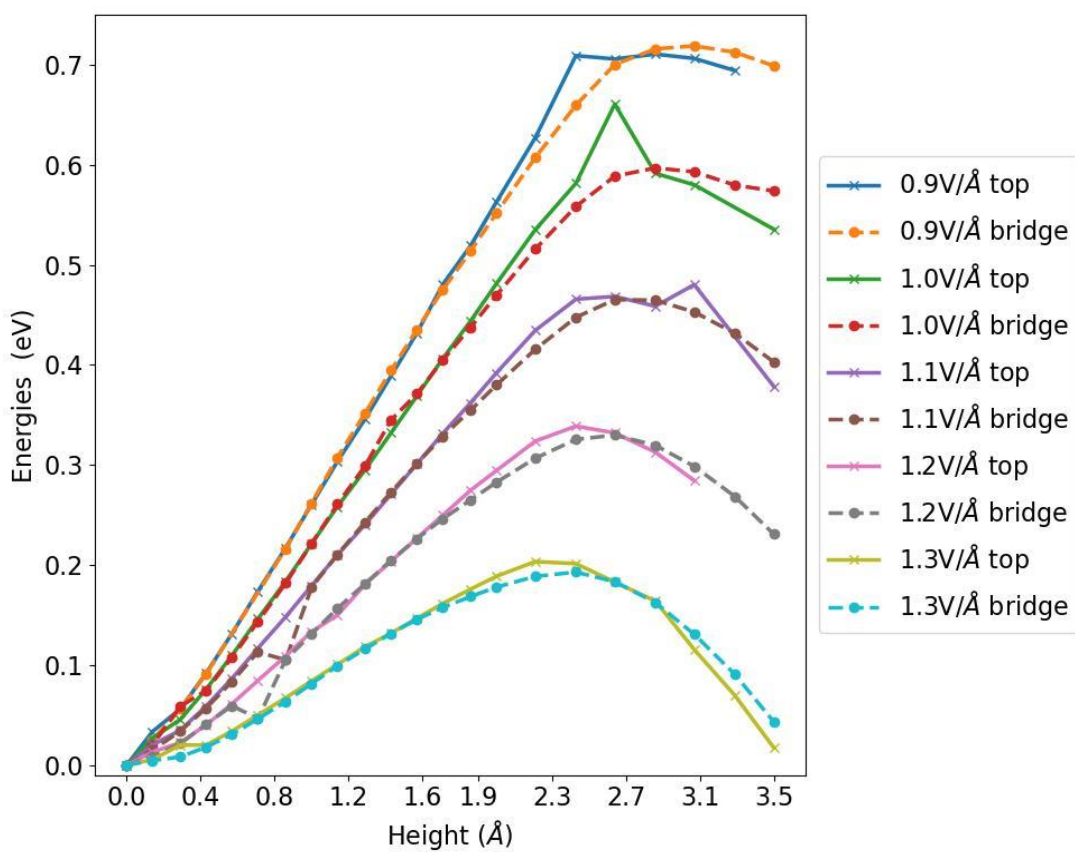

*Figure S. 1 The energy vs normal displacement curves at different electric fields for both on-top and bridge sites are shown. The presence of a barrier even at 1.3V/Å is visible for desorption from both the sites.*

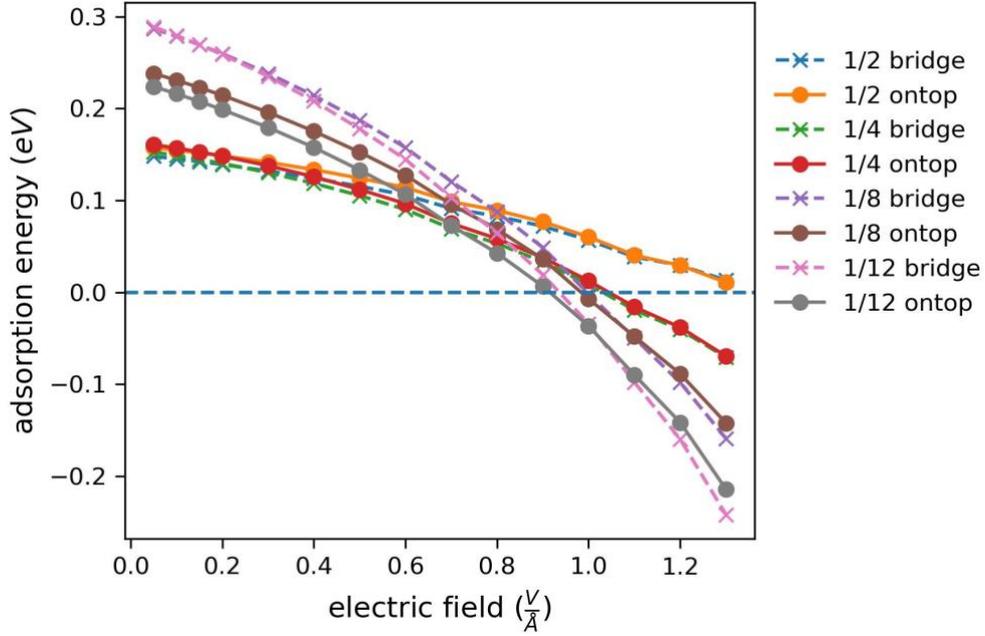

*Figure S. 2 Adsorption energies vs electric field for different coverages, (surface cells 1x1 (1/2), 1x2 (1/4), 2x2 (1/8), 3x2 (1/12)) of adatom (circles for on-top and crosses for bridge) are shown. At higher electric fields, the increasing coverage increases the adsorption energies.*

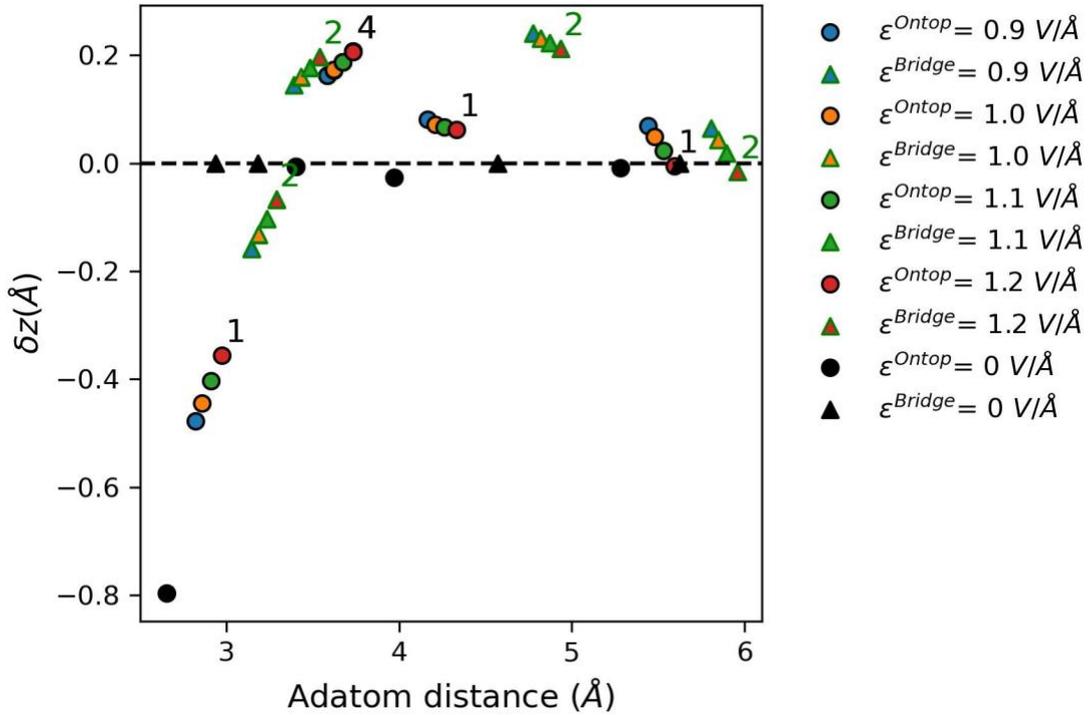

*Figure S. 3 The z relaxations of nearest neighbors of adatom (circles for on-top and triangles for bridge) at different electric fields are plotted as a function of their distance from the adatom. The annotated numbers indicate the number of nearest neighbors at that distance from the adatom.*